\documentclass[12pt, a4paper]{article}
\usepackage{amsfonts,amssymb,amsmath, amsthm, graphicx}

\begin{document}

\title{Casimir interaction: pistons and cavity.}

\author{Valery N. Marachevsky \thanks{email: maraval@mail.ru} \\
{\it Laboratoire Kastler Brossel, CNRS, ENS, UPMC,} \\
{\it Campus Jussieu case 74, 75252 Paris, France} \\{\it V. A. Fock
Institute of Physics, St. Petersburg
State University,}\\
{\it 198504 St. Petersburg, Russia} }



\maketitle

\begin{abstract}
The energy of a perfectly conducting rectangular cavity is studied
by making use of pistons' interactions. The exact solution for a 3D
perfectly conducting piston with an arbitrary cross section is being
discussed.
\end{abstract}

\section{Introduction}
The Casimir effect \cite{Casimir} was studied in various specific
cases and geometries \cite{Bordag1}. A new geometry that recently
attracted attention in the theory of the Casimir effect is the
piston geometry.

A piston plate is perpendicular to the walls of  a semi-infinite
cylinder and moves freely inside it, this geometry was first
investigated in a $2{\rm D}$ Dirichlet model \cite{Cavalcanti}.

An exact solution for a perfectly conducting square piston at zero
temperature was found in a $3{\rm D}$ model in the electromagnetic
and scalar cases \cite{Jaffe2} by making use of a geometric optics
approach; the limit of short distances between the piston and the
base of a cylinder was found in \cite{Jaffe2, Jaffe3} for an
arbitrary cross section of a piston; rectangular geometries and
finite temperatures were considered in \cite{Jaffe3}.

In this paper and our previous related papers \cite{Mar1, Mar2,
Mar3} we considered a slightly different geometry  - two piston
plates  inside an infinite cylinder, which yielded the same results
for rectangular pistons as in the case of a semi-infinite cylinder
due to perfectly conducting boundary conditions. In \cite{Mar1,
Mar2, Mar3} an exact solution for arbitrary cross sections and
arbitrary distances between piston plates was found at zero and
finite temperatures in the electromagnetic $3{\rm D}$ case.
Rectangular and circular cross sections are special cases of our
general solution.

A dilute circular piston and cylinder were studied perturbatively in
\cite{Barton}. In this case the force on two plates inside a
cylinder and the force in a piston geometry differ essentially. The
force in a piston geometry can change sign in this approximation for
thin enough walls of the material.

Different examples of pistons in a scalar case were investigated in
\cite{Fulling, Edery, Zhai}.

The case when the piston's cross section differs from that of a
cylinder was recently studied numerically \cite{Rod} and by means of
a geometric optics approach \cite{Jaffe4}.

Everywhere in this paper we study Casimir energies of an
electromagnetic field with perfectly conducting boundary conditions
imposed. First we study the energy of a rectangular cavity by making
use of pistons' interactions. Then we generalize the formulas for
the case of a 3D piston with an arbitrary cross section and consider
several special and limiting cases. Our formulas can be applied in
every case when the two dimensional Dirichlet and Neumann boundary
problems for Helmholtz equation can be solved analytically or
numerically.

We take $\hbar=c=1$.

\section{Construction of a rectangular cavity}
The Casimir energy can be regularized as follows:
\begin{equation} E = \frac{1}{2} \sum_{\omega_l} \omega_l^{-s} ,
\label{f23}
\end{equation}
where $s$ is large enough to make (\ref{f23}) convergent. Then it
should be continued analytically (\ref{f23}) to the value $s=-1$ ,
this procedure yields the renormalized finite Casimir energy. The
regularized electromagnetic Casimir energy for the rectangular
cavity $E_{cavity}(a,b,c,s)$ can be written in terms of Epstein $Z_3
\bigl(\frac{1}{a}, \frac{1}{b}, \frac{1}{c} ; s \bigr)$ and Riemann
$\zeta_R(s)$ zeta functions:
\begin{equation}
\begin{split}
E_{cavity}(a,b,c,s) &= \frac{\pi}{8} \Biggl(
\sum_{n_1,n_2,n_3=-\infty}^{+\infty\:\:\prime}
\Bigl[\Bigl(\frac{n_1}{a}\Bigr)^2 + \Bigl(\frac{n_2}{b}\Bigr)^2 +
\Bigl(\frac{n_3}{c}\Bigr)^2 \Bigr]^{-s/2} - \\
& \quad - 2 \Bigl(\frac{1}{a} + \frac{1}{b} + \frac{1}{c} \Bigr)
\sum_{n=1}^{+\infty} \frac{1}{n^s} \Biggr)
\end{split}
\end{equation}
\begin{align}
 Z_3 \Bigl(\frac{1}{a},
\frac{1}{b}, \frac{1}{c} ; s \Bigr) &=
\sum_{n_1,n_2,n_3=-\infty}^{+\infty\:\:\prime}
\Bigl[\Bigl(\frac{n_1}{a}\Bigr)^2 + \Bigl(\frac{n_2}{b}\Bigr)^2 +
\Bigl(\frac{n_3}{c}\Bigr)^2 \Bigr]^{-s/2} \\
\zeta_R (s) &= \sum_{n=1}^{+\infty} \frac{1}{n^s}
\end{align}
 The prime means that the term with all $n_i = 0$ should be
excluded from the sum. The reflection formulas for an analytical
continuation of zeta functions exist:
\begin{align}
&\Gamma \Bigl(\frac{s}{2}\Bigr) \pi^{-s/2} \zeta_R (s) = \Gamma
\Bigl(\frac{1-s}{2}\Bigr) \pi^{(s-1)/2} \zeta_R (1-s) \label{refl1}
\\ &\Gamma \Bigl(\frac{s}{2}\Bigr) \pi^{-s/2} Z_3 (a,b,c;s) =
(abc)^{-1} \Gamma \Bigl(\frac{3-s}{2}\Bigr) \pi^{(s-3)/2} Z_3
\Bigl(\frac{1}{a},\frac{1}{b},\frac{1}{c}; 3-s \Bigr) .\label{refl2}
\end{align}

The renormalized Casimir energy for a perfectly conducting
rectangular cavity can therefore be written as \cite{Lukosz}:
\begin{equation}
E_{cavity} (a,b,c) = - \frac{abc}{16\pi^2} Z_3(a,b,c; 4) +
\frac{\pi}{48} \Bigl(\frac{1}{a} + \frac{1}{b} + \frac{1}{c} \Bigr)
. \label{f24}
\end{equation}
The expression (\ref{f24}) can be rewritten in a different
mathematical form \cite{Mar1} :
\begin{multline}
E_{cavity} (a,b,c) =  \frac{\pi}{48 a} + \frac{\pi}{48 b} +
\frac{\pi}{48 c} + a E_{waveguide}(b,c) + \\ + \frac{1}{4}
\sum_{n_2, n_3 = -\infty}^{+\infty} \int_{-\infty}^{+\infty}
\frac{dp}{2\pi} \ln \biggl(1 - \exp \Bigl[-2 a \sqrt{\Bigl(\frac{\pi
n_2}{b}\Bigr)^2 + \Bigl(\frac{\pi n_3}{c}\Bigr)^2 + p^2}
\Bigr]\biggr) . \label{box}
\end{multline}
Here the Casimir energy for a unit length of a rectangular waveguide
is ($t=b/c$ or $t=c/b$) :
\begin{align}
 &E_{waveguide} (b, c) = E_{waveguide} (c, b) = \nonumber \\ &= -\frac{\pi^2}{720 t^2 bc}
+ \frac{t}{4\pi bc}
 \sum_{n=-\infty}^{+\infty} \int_{0}^{+\infty} dp \,\, p \,\,
 \ln \biggl( 1 - \exp\Bigl(-2\sqrt{\frac{\pi^2 n^2}{t^2} + p^2 }
 \Bigr) \biggr)= \nonumber \\
 &= -\frac{\pi^2}{720 t^2 bc} - \frac{1}{bc}\sum_{n=1}^{+\infty}
 \biggl(\frac{\csc^2{(n\pi/t )}}{16 n^2} + \frac{t \coth{(n\pi/t)}}{16\pi n^3} \biggr)
  \label{wave}
\end{align}

Let us discuss different terms appearing in (\ref{box})
 and thus clarify the physical meaning of a zeta function regularization
 in this case.

\begin{figure}
\centering \includegraphics[width=12cm]{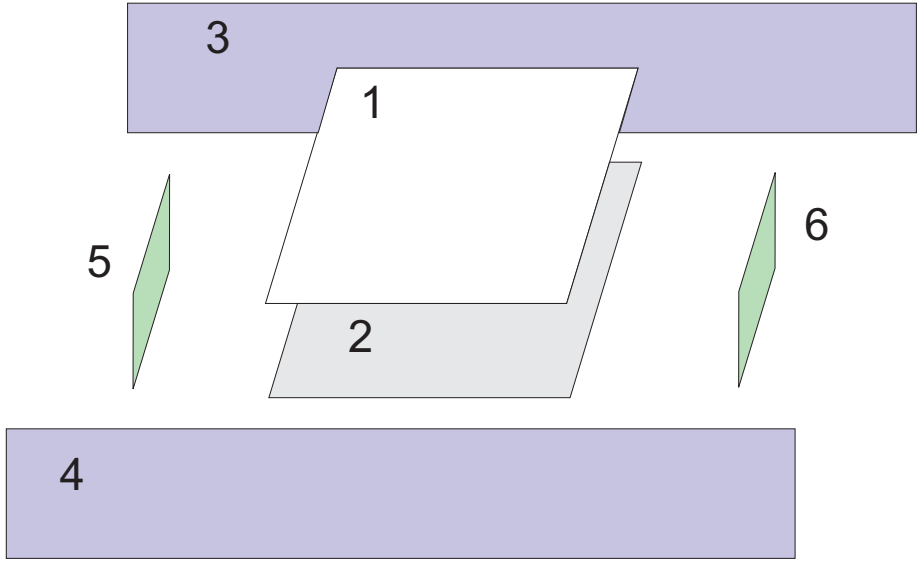}
\caption{Construction of a cavity}
\end{figure}

  Imagine that a piston is large in
two dimensions with sides $a_0$ and $c_0$ (plates $1$ and $2$ on
Fig.$1$). In this case the contribution to the energy in this
geometry is given by the Casimir result for two parallel plates with
the surface area $a_0 c_0$ and the edge term $\pi/48 b$:
\begin{equation}
-\frac{\pi^2 a_0 c_0}{720 b^3} + \frac{\pi}{48 b} , \label{ch1}
\end{equation}
where one of the terms in $a_0 E_{waveguide} (b, c_0)$ ($t=b/c_0$)
is taken into account.

The next step is to move other pistons (plates $3$  and $4$ on
Fig.$1$) that have a large side $a_0$ between these already existing
parallel plates. The energy change is equal to
\begin{equation}
a_0 E_{3-4}(b,c) + \frac{\pi}{48 c} , \label{ch2}
\end{equation}
where
\begin{equation}
E_{3-4}(b,c) = - \frac{1}{bc}\sum_{n=1}^{+\infty}
 \biggl(\frac{\csc^2{(n\pi/t )}}{16 n^2} + \frac{t \coth{(n\pi/t)}}{16\pi n^3} \biggr)
 .
  \label{lm1}
\end{equation}
In (\ref{ch2}) another term in $a_0 E_{waveguide} (b, c)$ was taken
into account ($c$ is a distance between the two piston plates $3$
and $4$, $t=b/c=c/b$). From the energy change (\ref{ch2}) it is
straightforward to obtain the force on the piston plates $3$ and
$4$, and in the limit $a_0 \to \infty$   one immediately obtains the
exact force on a unit length of stripes $3$ and $4$ from
$E_{3-4}(b,c)$.

The expression for the energy change (\ref{ch1})  is valid only when
$b$ is much smaller than the sizes of the plates $1$,$2$, and
(\ref{ch2}) is valid when $c$ is much smaller than $a_0$ and of the
order $b$ or less.

Some comments are needed to clarify the meaning of the terms
$\pi/48b + \pi/48c$ in the energy expression (\ref{box}). These
terms appear due to edges of the piston. They are precisely equal to
next to leading order terms in the expansion (\ref{r23}), which
means that they account for $4$ rectangular edges ($\chi=1/4$) of
the finite piston in two different expansions (for small $b$ and
small $c$).

The next possible step is to insert pistons $5$ and $6$ from the
opposite sides of the existing waveguide with sides $b$ and $c$ and
move them towards each other. The term
\begin{align}
&E_{5-6}(a,b,c) = \nonumber \\ &= \frac{1}{4} \sum_{n_2, n_3 =
-\infty}^{+\infty \;\;\;\:\prime} \int_{-\infty}^{+\infty}
\frac{dp}{2\pi} \ln \biggl(1 - \exp \Bigl[-2 a \sqrt{\Bigl(\frac{\pi
n_2}{b}\Bigr)^2 + \Bigl(\frac{\pi n_3}{c}\Bigr)^2 + p^2}
\Bigr]\biggr)  \label{lm3}
\end{align}
 yields the interaction energy of two plates distance $a$ apart inside
 an infinite  rectangular cylinder with
sides $b$ and $c$ (the term $n2=n3=0$ is excluded from the sum due
to a cancellation with the term $\pi/(48a)$ in (\ref{box})). The
force on the pistons $5$ and $6$ is straightforward.


In summary, the expression
\begin{equation}
\Delta E = E_{cavity}(a,b,c) - \frac{\pi}{48 b} - \frac{\pi}{48 c} +
\frac{ac}{a_0 c_0}\frac{\pi}{48 b} + \frac{a}{a_0} \frac{\pi}{48 c}
\label{E20}
\end{equation}
yields the energy change inside a cavity volume with sides $(a,b,c)$
during a construction of the following system:
\begin{description}
\item[Step $1$:]    pistons  $1$ and $2$ are being moved inside a
waveguide with large sides $a_0, c_0$ from a large distance between
them towards each other until the distance $b$ between them is
achieved. For a validity of the energy change (\ref{ch1})
 it is necessary to assume $b \ll a_0, b \ll c_0$.
\item[Step $2$:]   pistons $3$ and $4$  with sides $a_0$, $b$ are being moved
inside the existing box $(a_0, b, c_0)$ between the existing pistons
$1$ and $2$ towards each other until the distance $c$ between them
is achieved. For a validity of the energy change (\ref{ch2}) it is
necessary to assume $c \ll a_0$, also $c \ll c_0$.
\item[Step $3$:] pistons $5$ and $6$ with sides $b, c$ are being moved inside
the existing box $(a_0, b, c)$ towards each other until the distance
$a$ between them is achieved. It is assumed here that $a\ll a_0$.
However, the formula (\ref{lm3}) itself is exact for arbitrary
values of $a$ in the limit $a_0 \to \infty$, i.e. for plates inside
an infinite waveguide.
\end{description}

During each step the energy in the system decreases.

In the limit of infinite plates $1,2$ and stripes $3,4$ ($a_0,
c_0\to\infty$) the total energy change (steps $1-3$) inside the
cavity volume with sides $(a,b,c)$ can be written in the form:
\begin{equation}
\Delta E_{tot} = - \frac{abc}{16\pi^2} Z_3(a,b,c; 4) +
\frac{\pi}{48a}  \label{a30}
\end{equation}



\section{Arbitrary cross section results}

Suppose there are two plates inside an infinite cylinder of an
arbitrary cross section $M$ (Fig.$2$). To calculate the force on
each plate imagine that $4$ parallel plates are inserted inside an
infinite cylinder and then $2$ exterior plates are moved to spatial
infinity. This situation is exactly equivalent to $3$ perfectly
conducting cavities touching each other. From the energy of this
system one has to subtract the Casimir energy of an infinite
cylinder without plates inside it. Doing so we obtain the energy of
interaction between the interior parallel plates and the attractive
force on each interior plate inside the cylinder:
\begin{align}
\mathcal{E} (a) &= \sum_{\omega_{wave}} \frac{1}{2} \ln (1-\exp(-2 a
\, \omega_{wave}))   \label{r7}\\
F (a) &= - \frac{\partial \mathcal{E}(a)}{\partial a}, \label{a13}
\end{align}
the sum here is over all  $TE$ and $TM$ eigenfrequencies
$\omega_{wave}$ for a cylinder with the cross section $M$ and an
infinite length.

In fact, the Casimir energy of our electromagnetic system is
proportional to the sum of free energies for two boson scalar fields
(with Dirichlet and Neumann boundary conditions imposed  at the
boundary of an infinite cylinder with a cross section $M$ and zero
Neumann eigenvalue excluded) at finite temperature $T=1/\beta$ if we
make a substitution $a \to \beta/2$. Free energies have a well
defined finite part, their sum up to a factor $1/a$ coincides with
(\ref{r7}).

 One can rewrite (\ref{r7}) in a different form:
 \begin{equation}
 \mathcal{E}(a) = -\frac{1}{2\pi} \sum_{l=1}^{+\infty} \biggl(\sum_{\lambda_{k D}}
 \frac{\lambda_{k D} K_1 (2 l\lambda_{k D} a)}{l}
 + \sum_{\lambda_{i N}}
  \frac{\lambda_{i N} K_1 (2 l\lambda_{i N} a)}{l}\biggr)
  .\label{p10}
 \end{equation}
Here
\begin{align}
& \Delta^{(2)} f_k (x,y) = - \lambda_{k D}^2 f_k (x,y)  \label{Helm1} \\
&  f_k (x,y) |_{\partial M} = 0 \nonumber \\
& \Delta^{(2)} g_i (x,y) = - \lambda_{i N}^2 g_i (x,y)  \label{Helm4} \\
& \frac{\partial g_i (x,y)}{\partial n} \Bigl|_{\partial M} = 0 .
\nonumber
\end{align}
The results (\ref{r7}), (\ref{p10}) are our main zero temperature
results. Our results are {\it exact for an arbitrary curved geometry
of a cylinder}.

\begin{figure}
\centering \includegraphics[width=10cm]{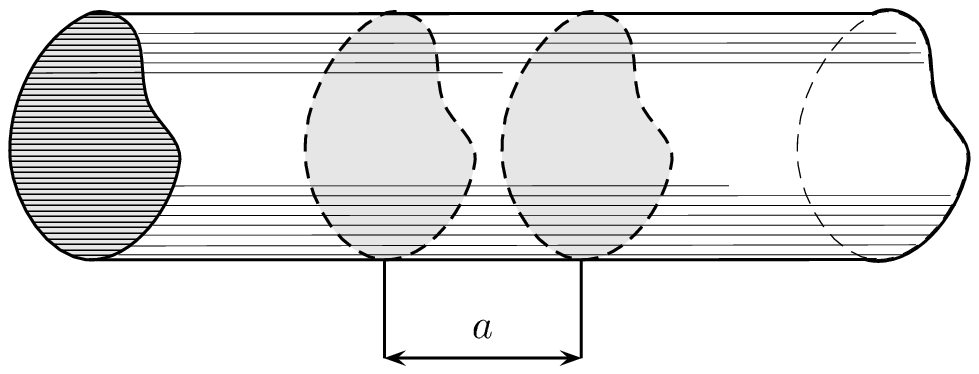} \caption{Two
plates inside an infinite cylinder}
\end{figure}

For a rectangular cylinder with sides $b$ and $c$ the exact Casimir
energy of two plates inside it can be written as:
\begin{equation}
\mathcal{E}_{rect} (a) =
-\sum_{l=1}^{+\infty}\sum_{m,n=-\infty}^{+\infty \: \prime}
\frac{\sqrt{m^2/b^2 + n^2/c^2}}{4l} \:\:  K_{1} (2 l \pi a
\sqrt{m^2/b^2 + n^2/c^2}) . \label{p11}
\end{equation}
The term $m=n=0$ is omitted in the sum.

For a circular cylinder the eigenvalues of the two dimensional
Helmholtz equation $\lambda_{k D} , \lambda_{i N}$ are determined by
the roots of Bessel functions and derivatives of Bessel functions.
The exact Casimir energy of two circular plates of the radius $b$
separated by a distance $a$ inside an infinite circular cylinder of
the radius $b$ is given by:
\begin{align}
&\mathcal{E}_{circ}(a) =  - \sum_{l=1}^{+\infty} \sum_{\nu =
0}^{+\infty} \sum_{j} \frac{1}{2\pi b}\frac{\mu_{D \nu j} K_1 (2 l
\mu_{D \nu j} a/b )
 +  \mu_{N \nu j} K_1 (2 l \mu_{N \nu j} a/b)}{l} ,\label{p12}\\
    &J_{\nu}(\mu_{D \nu j}) = 0 ,
    \:\:\:\: J_{\nu}^{'} (\mu_{N \nu j}) = 0   . \nonumber
 \end{align}
The sum is over positive $\mu_{D \nu j}$ and $\mu_{N \nu j}$.

The leading asymptotic behaviour of $\mathcal{E}(a)$ for long
distances $\lambda_{1 D} a \gg 1$, $\lambda_{1 N} a \gg 1$ is
determined by the lowest positive eigenvalues of the two dimensional
Dirichlet and Neumann problems $\lambda_{1 D}, \lambda_{1 N} $:
\begin{equation}
\mathcal{E}(a)|_{\lambda_{1 D} a \gg 1, \: \lambda_{1 N} a \gg 1 }
\sim - \frac{1}{4\sqrt{\pi a}} \Bigl( \sqrt{\lambda_{1D}} e^{-2
\lambda_{1D} a} + \sqrt{\lambda_{1N}} e^{-2 \lambda_{1N} a} \Bigr) ,
\label{om1}
\end{equation}
so the Casimir force between the two plates in a cylinder is
exponentially small for long distances. This important property of
the solution is due to the gap in the frequency spectrum or, in
other words, it is due to the finite size of the cross section of
the cylinder. Due to this property one needs a finite number of the
eigenvalues of  the Helmholtz equation for $2D$ Dirichlet and
Neumann boundary problems (\ref{Helm1}-\ref{Helm4}) to obtain the
Casimir energy at a specific distance $a$ between the plates with a
desired accuracy.

The free energy at a temperature $T=1/\beta$ describing the
interaction of two parallel perfectly conducting plates inside an
infinite perfectly conducting cylinder with the cross section $M$
has the form:
\begin{align}
& \mathcal{F} (a,\beta) = \nonumber \\ & =\frac{1}{\beta}
\sum_{\lambda_{k D}} \sum_{m=-\infty}^{+\infty} \: \frac{1}{2} \ln
\Bigl(1-\exp
\bigl(-2a\sqrt{\lambda_{k D}^2 + p_m^2} \bigr) \Bigr) + \nonumber \\
&+ \frac{1}{\beta} \sum_{\lambda_{i N}} \sum_{m=-\infty}^{+\infty}
\: \frac{1}{2} \ln \Bigl(1-\exp \bigl(-2a\sqrt{\lambda_{i N}^2 +
p_m^2} \bigr) \Bigr) . \label{r21}
\end{align}
This is our central finite temperature result. Note that
$\lambda_{iN} \ne 0$. We used the standard notation $p_m = 2 \pi m
T$.

In the long distance limit $a \gg \beta/(4\pi)$ one has to keep only
$m=0$ term in $(\ref{r21})$. Thus the free energy of the plates
inside a cylinder in the  high temperature limit is equal to:
\begin{eqnarray}
\mathcal{F}(a, \beta)|_{a \gg \beta/(4\pi)} = & \frac{1}{2\beta}
\sum_{\lambda_{k D}}  \: \ln \Bigl(1-\exp (-2a\lambda_{k D} ) \Bigr)
+ \nonumber \\ & \frac{1}{2\beta} \sum_{\lambda_{i N}} \: \ln
\Bigl(1-\exp (-2a\lambda_{i N} ) \Bigr) .\label{r24}
\end{eqnarray}
One can check that the limit $\lambda_{1 D} a \ll 1$, $\lambda_{1 N}
a \ll 1$ in (\ref{r24}) immediately yields the known high
temperature result for two parallel perfectly conducting plates
separated by a distance $a$ \cite{Brevik}.

For $a \ll \beta/(4\pi)$ and $\lambda_{1 D} a \ll 1$, $\lambda_{1 N}
a \ll 1$ one can use the heat kernel expansion \cite{Vassilevich,
Gil} and properties of the zeta function \cite{Kirsten, Santangelo2,
Elizalde} to obtain the leading terms for the free energy:
\begin{equation}
\mathcal{F}(a,\beta)|_{a \ll \beta/(4\pi), \: \lambda_{1 D} a \ll 1,
\: \lambda_{1 N} a \ll 1} = -\frac{\zeta_R(4)}{8\pi^2}\frac{S}{a^3}
+ \frac{\zeta_R (2)}{4\pi a} (1 - 2 \chi) + O(1) , \label{r23}
\end{equation}
where
\begin{equation}
\chi = \sum_i \frac{1}{24} \Bigl(\frac{\pi}{\alpha_i}-
\frac{\alpha_i}{\pi}\Bigr) + \sum_j \frac{1}{12\pi} \int_{\gamma_j}
L_{aa} (\gamma_j) d\gamma_j.
\end{equation}
Here $S$ is an area of the cross section $M$, $\alpha_i$ is the
interior angle of each sharp corner at the boundary $\partial M$ and
$L_{aa} (\gamma_j)$ is the curvature of each boundary smooth section
described by the curve $\gamma_j$. The force calculated from
(\ref{r23}) coincides with the zero temperature force $F_C$ in
\cite{Jaffe2}, (Eq.7).

\section*{Acknowledgments}
 It is a pleasure to thank Michael Bordag for hospitality
 in Leipzig during ${\rm QFEXT-07}$, also all colleagues for fruitful
 discussions in Leipzig. This work has been supported
 by a CNRS grant ANR-06-NANO-062 and
  grants RNP $2.1.1.1112$, SS $.5538.2006.2$, RFBR $07-01-00692-a$.


\end{document}